


\input harvmac
\ifx\epsfbox\UnDeFiNeD\message{(NO epsf.tex, FIGURES WILL BE IGNORED)}
\def\figin#1{\vskip 0.5in}
\else\message{(FIGURES WILL BE INCLUDED)}\def\figin#1{#1}\fi
\def\ifig#1#2#3{\xdef#1{fig.~\the\figno}
\goodbreak\midinsert\figin{\centerline{#3}}%
\vskip -0.5in\centerline{\vbox{\baselineskip12pt
\advance\hsize by -1truein\noindent\footnotefont{\bf Fig.~\the\figno:} #2}}
\bigskip\endinsert\global\advance\figno by1}

\Title{\vbox{\hbox{CU-TP-620}\hbox{hep-th/9312149}}}{$S^1 \times S^2$
wormholes and topological
charge*}\footnote{}{* This work supported in part by the U.S.\ Department of
Energy}

\centerline{S. Alexander Ridgway$^\dagger$\footnote{}{$^\dagger$
{\it E-mail:} {\tt ridgway@cuphyf.phys.columbia.edu}}}
\bigskip
\centerline{Physics Department}
\centerline{Columbia University}
\centerline{New York, New York \ 10027}
\vskip .3in
I investigate solutions
to the Euclidean Einstein-matter field equations with topology
$S^1 \times S^2 \times R$ in a theory with
a massless periodic scalar field and electromagnetism.  These solutions
carry winding number of the periodic scalar as well as magnetic
flux.  They induce violations of a quasi-topological
conservation law which conserves the product of magnetic flux and
winding number on the background spacetime. I extend these solutions
to a model with stable loops of superconducting cosmic string, and
interpret them as contributing to the decay of such loops.

\Date{December 1993}

\def\to{\!\rightarrow\!}
\def\frac#1#2{{\textstyle {#1 \over #2}}}

\def\pmb#1{\setbox0=\hbox{#1}\kern-.01em\copy0\kern-\wd0
  \kern.02em\copy0\kern-\wd0\kern-.01em\raise.017em\box0}

\def\d{{\rm d}}
\def\half{\frac{1}{2}}

\def\intst{\int\!{\rm d}^4 x \sqrt{g} }
\def\pl#1{{\it Phys.~Lett.} {\bf #1}}
\def\nuc#1{{\it Nucl.~Phys.} {\bf #1}}
\def\pr#1{{\it Phys.~Rev.} {\bf #1}}

\def\prl#1{{\it Phys.~Rev.~Lett.} {\bf #1}}

\baselineskip=17pt plus 2pt minus 1pt
\newsec{Introduction}

One of the factors driving the development of the wormhole formalism has
been the existence of classical wormhole solutions to the Euclidean
equations of motion for gravity and matter.  A wormhole solution is
typically defined as a solution that is asymptotic to two distinct flat
spacetimes.  A configuration of wormhole topology that is not a solution to
the equations of motion is less convincing evidence of topology change than is
a classical solution, as the classical solution may give the
dominant contribution to some quantum-mechanical
amplitude in the semiclassical limit.  Since this is
why one considers classical solutions, it is reasonable to ask
to which quantum-mechanical amplitudes a proposed wormhole solution will give
the dominant contribution.

One could propose that the dominant contribution is to an amplitude for the
creation or annihilation of a baby universe. Since this process is expected
to be unobservable for low-energy processes on a background spacetime, one
suspects that this is not a measurable amplitude, and therefore not a
physically relevant calculation.
One can instead require, however, that the
wormhole mediate some process that otherwise could not take place in the
underlying field theory.  In a field theory with some conserved global
charge, charge violation is just such a process.

The first wormhole solutions found by Giddings and
Strominger\ref\gands{S.~Giddings and A.~Strominger, \nuc{B306}:890 (1988)}
carry flux associated with a three-index antisymmetric tensor field, or
axion.  These wormholes thus violate axion charge conservation.  Lee
showed\ref\klee{K.~Lee, \prl{61}:263 (1988)} how to represent this in terms
of a massless scalar field
dual to the three-index tensor,
$$ H_{\mu\nu\lambda} = \epsilon_{\mu\nu\lambda\sigma} \partial^\sigma\! a .$$
The effect of wormholes can then be represented by operators that explicitly
break
the symmetry $a \to a + c$.  In order to find the solutions for the scalar
field, or for the more general case of a complex scalar field with a $U(1)$
global symmetry,\ref\claw{S.~Coleman and K.~Lee, \nuc{B329}:387 (1990)\semi
L.~F.~Abbott and M.~B.~Wise, \nuc{B325}:687 (1989)} one must constrain
the initial and final states
to be states of definite charge.  This can result in certain terms in the
equations of motion
changing signs, the net result being that the solutions for the
three-index tensor theory are identical to the solutions for the dual
scalar field theory, which would not be the case if the equations of
motion were applied naively.

Wormhole solutions have also been found in
$3$-dimensional\foot{``$n$-dimensional'' will refer to
$n$-dimensional Euclidean spacetime,
with $n-1$ space dimensions and one Euclidean time dimension.}
electromagnetism by Hosoya and Ogura;\ref\hosogu{A.~Hosoya and W.~Ogura,
\pl{B225}:117 (1989)} these solutions carry
magnetic flux down the wormhole throat.  These solutions are really the
direct $3$-dimensional analog of the $4$-dimensional Giddings-Strominger
axionic wormholes.  In both cases, the charge that supports the wormhole
throat is topologically conserved.  This means that the current
conservation equation is an identity when expressed in terms of the gauge
potential.  In three dimensions, the magnetic flux current is
$$ j^\lambda = \epsilon^{\lambda\mu\nu} F_{\mu\nu} . $$
The flux conservation equation, $\partial_\mu j^\mu = 0$, is an identity
when $F$ is written in terms of the gauge potential $A_\mu$.

The effects of these
wormholes\ref\magworm{A.~K.~Gupta, J.~Hughes, J.~Preskill and M.~B.~Wise,
\nuc{B333}:195 (1990)} will be similar to the effects
of any finite-action monopole solutions that may exist in the theory.
In a three-dimensional theory, a
monopole solution can be thought of as an instanton that mediates processes
that violate magnetic flux conservation.  One can express the effect of such
violation in terms of a scalar field dual to $F$, just as in the
four-dimensional case one can express the effects of axion charge
non-conservation in terms of a scalar field dual to $H$.

In this work, I generalize the magnetic wormhole to
four-di\-men\-sion\-al electromagnetism.  In four dimensions, one
still has magnetic flux conservation, in that magnetic flux lines cannot
end. (On a spatial slice, ${\pmb{\hbox{$ \nabla $}}} \cdot {\bf B} = 0$.)
Loops of
magnetic flux can, however, shrink to nothing.  This can be avoided
by giving the wormholes the topology $S^2 \times S^1
\times R$, so that magnetic flux on the two-sphere can wind around the
circle.  I will also put a topologically conserved charge on $S^1$,
the winding number for a periodic scalar field.\foot{Keay
and Laflamme\ref\keaylaflamme{B.~J.~Keay and R.~Laflamme, \pr{D40}:2118
(1989)} have also constructed wormhole solutions
of the same topology in the dual of this theory. These
wormholes have axion charge rather than winding number.}

I will then discuss the effects of such wormholes.  I believe that while
insertions of the usual $S^3 \times R$ wormholes induce pointlike
operators at low energy, $S^1 \times S^2 \times R$ wormholes
induce looplike operators.  I will discuss the consequences of
this, in particular in a model with stable loops of superconducting
cosmic string.

\newsec{Solutions}

I construct wormhole solutions in a theory that includes the electromagnetic
field and a massless periodic
scalar field (the axion) coupled to gravity.  These solutions will have
topology $S^1 \times S^2 \times R$. The
periodic scalar has a topological charge associated with it, the winding number
on $S^1$.  For the electromagnetic field, magnetic flux on the
two-sphere is conserved; this conservation law is topological both in the
sense that the current conservation law is an identity, and in the sense
that the flux is a topological invariant of the two-sphere if charged fields
are added to the theory.

The Euclidean action for this theory is given by
$$ S = \intst \left[ -{1 \over 16 \pi G} R + {v^2 \over 2} g^{\mu\nu}
       \partial_\mu \Theta \partial_\nu \Theta  + {1\over 4 e^2} g^{\mu\nu}
       g^{\lambda\sigma} F_{\mu\lambda} F_{\nu\sigma} \right]. $$
Here $v$ is the axion symmetry-breaking scale; $\Theta$ is a periodic
scalar field with period $2 \pi$.
Now the simplest possible ansatz for a wormhole solution with the desired
features will have a Euclidean Kantowski-Sachs
geometry:\ref\ksachs{M.~A.~H.~McCallum, in {\it General Relativity: An
Einstein Centenary Survey},\hfil\break
ed.~S.~W.~Hawking and W.~Israel (Cambridge
University Press, Cambridge, 1979)}
each spatial slice
will be homogeneous and characterized by the radius of the two-sphere, the
radius of the circle, and the topological charges associated with each.
The metric for this is
$$ \d s ^2 = N^2(\tau) \d \tau^2 + a^2(\tau) \d l^2 + b^2(\tau) \d \Omega^2, $$
where $l$ is a periodic coordinate with period $1$, and $\d \Omega^2$ is the
solid angle element on $S^2$,
$\d \Omega^2 =\d \theta^2 + \sin^2\!\theta \,\d \phi^2$.
By reparameterizing $\tau$, the
lapse function $N^2(\tau)$ can be set to any strictly positive function;
here I set it to unity. The quantities $a$ and $b$ are the radii of
the circle and the two-sphere, respectively.

The field equation for $\Theta$ is
$$ \partial_\mu [ \sqrt{g} g^{\mu\nu} \partial_\nu \Theta ] = 0. $$
I impose the restriction that $\Theta$ is a function of $l$ and $\tau$
only, and that derivatives of $\Theta$ are functions of $\tau$ only:
$$\Theta = T_l(\tau) l + T_0(\tau) . $$
Since $\Theta$ must have an integral winding number on the circle, it must
satisfy the boundary conditions
$$ \Theta(l,\tau) = \Theta(l+1,\tau) - 2 \pi n , $$
and thus
$$ \Theta = 2 \pi n l + T_0(\tau) . $$
I am interested in the case $ n \ne 0 $, and in this case, $T_0(\tau)$ must
be constant to avoid off-diagonal terms in the energy-momentum tensor. I
eliminate constant $T_0$ by shifting $\Theta$, so
$$ \Theta = 2 \pi n l . $$

For the electromagnetic field there is a
field equation and a Bianchi identity:
$$ \partial_\mu [\sqrt{g} F^{\mu\nu}] = 0 \hbox{\quad and \quad}
   F_{\mu\nu,\rho} + F_{\nu\rho,\mu} + F_{\rho\mu,\nu} = 0 .$$
For a purely magnetic solution, $F^{0i} = 0$; therefore,
$$ \partial_\mu [\sqrt{g} F^{\mu\nu}] =\partial_0 [\sqrt{g} F^{0\nu}]=0. $$
Thus the field equation is automatically satisfied.
For a homogeneous magnetic field on the two-sphere, the correct ansatz
(in the coordinate basis) is
$$ F_{\theta\phi} = - F_{\phi\theta} = {\Phi(\tau) \over 4 \pi} \sin\theta .$$
The Bianchi identity gives $\partial_\tau F_{\theta\phi} = 0$, and thus
$\Phi(\tau) = \Phi_0$, a constant.  $\Phi_0$ is the conserved magnetic flux
on the two-sphere.

The conservation laws allow one to solve the matter field equations quite
directly; the only non-trivial equations are Einstein's equations, which
are
$$ R_{\mu\nu} - \half g_{\mu\nu} R = 8\pi G T_{\mu\nu} . $$
For the given field content, the energy-momentum tensor is
$$ T_{\mu\nu} = v^2 [ \partial_\mu \Theta \partial_\nu \Theta - \half
    g_{\mu\nu}g^{\alpha\beta} \partial_\alpha \Theta \partial_\beta \Theta ]
  + {1\over e^2} [g^{\alpha\beta} F_{\alpha\mu} F_{\beta\nu} - \frac{1}{4}
g_{\mu\nu}g^{\alpha\beta}g^{\lambda\sigma}F_{\alpha\lambda}F_{\beta\sigma}],$$
where I include the metric explicitly.  Substituting the
ansatz for the metric and the solutions to the matter equations, Einstein's
equations reduce to the following three equations:
\def\ad{{\dot a}}
\def\add{{\ddot a}}
\def\bd{{\dot b}}
\def\bdd{{\ddot b}}
\eqna\einstein
$$ \eqalignno{
   {2 \ad \bd \over a b} + {\bd^2 \over b^2} - {1\over b^2} &=
   -{Q_1^2\over a^2} - {Q_2^2\over b^4}, & \einstein a \cr
   {2 \bdd \over b} + {\bd^2 \over b^2} - {1\over b^2} &=
   {Q_1^2\over a^2} - {Q_2^2\over b^4}, & \einstein b \cr
   {\add \over a} + {\bdd \over b} + {\ad \bd \over a b} &=
   -{Q_1^2\over a^2} + {Q_2^2\over b^4}, & \einstein c \cr} $$
where ${\dot {(\ )}} \equiv {\d \over \d\tau} (\ )$,
$Q_1^2 = 8 \pi G ( 2\pi^2 n^2 v^2)$ and $Q_2^2 = 8 \pi G \left({\Phi_0^2 \over
32 \pi^2 e^2} \right)$.
Equation \einstein{a}\ is a constraint equation for the system of
second-order differential equations defined by the other two; one can
easily verify that it is conserved by equations (b) and (c).
One can obtain either of equations (b) and (c) from equation (a) and the
remaining equation, thus
equations (b) and (c) are redundant and either can be eliminated without loss
of generality.

Do wormhole solutions to these equations exist?
What exactly do we mean by a wormhole solution?  Unlike
the case where the topology is $S^3 \times R$, there will {\it not} be
any asymptotically Euclidean solutions with topology $S^2\times S^1 \times
R$.  First, the ansatz itself
cannot be asymptotic to $R^4$, since the topology forbids it.  Second, while
there are configurations asymptotic to flat $R^3 \times S^1$
({\it i.e.,} $a \to \hbox{constant}$, $b \to \tau + \hbox{constant} $), a
simple
argument shows that these cannot be solutions when $Q_1 \ne 0$: Since
the circle goes to a constant radius, the energy density that is
due to the winding of the scalar goes
to a non-zero constant, and thus the curvature cannot go to zero as it
must for a flat solution.  So what {\it do} we mean by a wormhole?
In this case, I will define a wormhole solution to be
a solution such that a)  there exists a small ``throat'' where
both of the radii
attain a minimum value, and b) some distance outside the throat both radii
become much larger than they are near the throat and
are ``almost'' of the form
$a = \hbox{constant}$, $b = \tau + \hbox{constant}$. I will
demonstrate the existence of such solutions.

The first thing to determine is whether a wormhole throat can form.
The condition for a wormhole throat (at a particular value
of $\tau$) is that $\ad =  \bd = 0$ and $\add,\ \bdd  \ge 0$. Is this
consistent with the equations? Setting $\ad$ and $\bd$ to zero,
constraint equation~\einstein{a}\
gives relation between $a$ and $b$ at
the throat:
\eqn\throat{ {Q_1^2 \over a^2} = {1\over b^2} - {Q_2^2 \over b^4} .}
Using this relation and equation~\einstein{b}\ one has
$$ {\bdd \over b} = {Q_1^2 \over a^2} $$
at the throat, thus $\bdd$ will always be greater than zero. Finally, using
the previous two relations and \einstein{c}, at the throat,
$$ {\add \over a} = {3 Q_2^2 \over b^4} - {2 \over b^2} .$$
To make $\add$ positive at the throat requires $Q_2^2 / b^2 > 2/3$.
Note also that equation~\throat\ implicitly requires
that $Q_2^2 / b^2 < 1$ at the throat.
At the throat, then, $Q_2^2/b^2$ is a free parameter, which
must satisfy
\eqn\thrin{ 1 > {Q_2^2 \over b^2} > {2 \over 3}, }
and one can calculate all other quantities from this (and of course, from
$Q_1$ and $Q_2$).

First consider the case $Q_1 = 0$.  In Euclidean space,
the given ansatz is equivalent to static,
spherically symmetric spacetime with periodically identified time, where
the Euclidean time $\tau$ of the ansatz becomes the radial coordinate of the
spherically symmetric spacetime.
For $Q_1 = 0$, the solution to
these equations is well known: it is the Euclidean magnetic
Reissner-Nordstr\o m
solution.\ref\rn{S.~W.~Hawking, in {\it General Relativity: An
Einstein Centenary Survey},
ed.~S.~W.~Hawking and W.~Israel (Cambridge
University Press, Cambridge, 1979)} These solutions, of course, seem nothing
like wormholes, but there is, in fact, a solution with a ``throat'':
$$ \eqalign{ b &= b_0 \equiv {Q_2}, \cr
             a &= a_0 \cosh(\tau / b_0). \cr } $$
The constant radius of the two-sphere is equal to the horizon radius of the
extreme Reissner-Nordstr\o m black hole.  This solution is still nothing
like a wormhole; I will therefore ignore it and concentrate on $Q_1 > 0$.

In the general case, $Q_1, Q_2 > 0$, I was unable to obtain analytic
solutions.  I did obtain some results by numerically integrating the
system of ordinary differential equations~\einstein{}. One need
use only equations \einstein{a} and (b), which give a first-order
differential equation for $a(\tau)$ and a second-order differential
equation for $b(\tau)$:
\eqna\num
$$ \eqalignno{ {\d a\over \d \tau} &= {ab\over 2 \bd}
\left({1\over b^2} - {\bd^2\over b^2} - {Q_1^2\over a^2} -{Q_2^2\over b^4}
\right) & \num a \cr
{\d^2 b\over \d \tau^2} &= {b\over 2 }
\left({1\over b^2} - {\bd^2\over b^2} + {Q_1^2\over a^2} -{Q_2^2\over b^4}
\right). & \num b \cr } $$

I found numerical solutions by performing integrations with initial
conditions set to values appropriate for a wormhole throat:
$$ \eqalign{ b(0) &= b_0 \qquad\qquad
\left({2\over 3} < {Q_2^2\over b_0^2} < 1 \right) \cr
a(0) &= Q_1 b_0 \left(1 - {Q_2^2  \over b_0^2} \right)^{-1/2}\cr
\bd(0) &= \epsilon .\cr } $$
Note that $\bd(0)$ is set to a small, non-zero value $\epsilon$.  This is
necessary, because when $\bd = 0$, equation \einstein{a} merely imposes
the constraint \throat\ without fixing $\ad$. Setting $\bd$ to
$\epsilon$ at $\tau = 0$ does set $\ad(0)$ to (nearly) zero with $a(0)$ and
$b(0)$ as given.  I set $\epsilon$ small enough that computer test runs
with the opposite sign for $\epsilon$ showed no significant difference.
In general,
equation~\num{a} will be undefined whenever $\bd$ goes through
zero.  The integrator ({\it Mathematica} built-in) handled this without
difficulty. Graphs of some of the results are
displayed in
Figures 1--3.

What are some of the general features of the solutions that one can see
from the numerical results?  Note first that one only need calculate
solutions for fixed values of $Q_1$ and $Q_2$, here set to 1.
This is possible because if the pair of functions $a(\tau)$ and $b(\tau)$
is a solution to equations \einstein{}\ for any particular values of
$Q_1$ and $Q_2$, there is a corresponding solution with ${\tilde Q_1} =
\lambda_1 Q_1$, ${\tilde Q_2} = \lambda_2 Q_2$ given by
\eqn\scaling{
\eqalign{{\tilde a}(\tau) &= \lambda_1 \lambda_2 a(\tau/\lambda_2) \cr
{\tilde b}(\tau) &= \lambda_2 b(\tau/\lambda_2) . \cr} }
The qualitative features of the solutions thus depend only on the parameter
$Q_2^2/b_0^2$.

These scaling relations also yield the important relation between the
charge carried by the wormhole and its size.  For a fixed value of the
parameter $Q_2^2/b_0^2$, one can find a solution $A(\tau)$, $B(\tau)$,
for $Q_1 = Q_2 = 1$; then the general solution is
$$\eqalign{a(\tau) &= Q_1 Q_2 A(\tau/Q_2) \cr
           b(\tau) &= Q_2 B(\tau/Q_2) . \cr } $$
So the overall size of the solution is proportional to $Q_2$, while the
length of the $S^1$ loop is also proportional to $Q_1$.

I performed numerical integrations that
started at the wormhole throat with various values
for the free parameter $Q_2^2/b_0^2$.
When $Q_2^2/b_0^2$ is very close to 1, there is a wormhole-like  solution,
where $a$ starts to rise very rapidly for a time and then levels off,
seeming to approach a constant. The $S^2$ radius $b$ starts out fairly
flat, then goes to
a regime in which $\bd$ is nearly 1.  If one continues to integrate to much
larger values of $\tau$,
$a$ will reach a maximum and start decreasing,
eventually collapsing to zero, while $b$ increases rapidly to
infinity.  When $Q_2^2/b_0^2$ is not close to 1,
this ``nearly flat'' behavior never begins;
instead, $a$ just reaches a maximum and then collapses while $b$
diverges --- the only difference is that this happens much sooner,
never allowing the solution to reach a nearly flat regime.

\newsec{Wormhole Insertions and Topological Charge}

What relevance do these wormhole solutions have
for a theory of quantum gravity?  In particular,
what kind of effects does this type of wormhole have on low-energy
physics in flat, four-dimensional spacetime?  To determine
this, one must first understand how these wormhole geometries of
$S^1 \times S^2 \times R$ topology
can attach to flat $R^4$.

The answer to this question is actually suggested by the solutions
themselves. The metric given by $a(\tau) = \hbox{constant}$
and $b(\tau) = \tau$ is flat; flat $R^3 \times S^1$ has this metric with
$\tau \in [0,\infty)$.  The subset of this space given by
$\tau \in [0,\tau_f]$, is flat $B^3\times S^1$,
where $B^3$ is the three-dimensional ball.  There
are $B^3 \times S^1$ subsets of $R^4$ that consist of a loop in spacetime
with a neighborhood around it. The geometry of these subsets very nearly
approximates that of the flat $B^3 \times S^1$ described above, at least
in the limit where the loop (the $S^1$) is long and straight on the scale
of the ball
around it.  After excising such a region from the background, there is
a boundary left with topology $S^2 \times S^1$, to which one can attach
the $S^2 \times S^1$ boundary of flat $B^3\times S^1$,
or any other geometry that approximates this near the
boundary.  The given wormhole solutions almost match this geometry
in the $\ad \cong 0$, $\bd \cong 1$ regime.  It should then be possible
to perturb the
geometries of the wormhole solution and the background spacetime in such a
way that they can be patched together on the $S^1 \times S^2$ boundary.
The geometry formed this way is ``almost'' a classical solution.
I conjecture that in theories with appropriate matter content, there
exists an exact classical solution which closely approximates this
geometry.  It is this solution which one should think of as the
wormhole under discussion.

\def\C{{\cal C}}
What happens in the background of such a solution?
{}From the point of view of the background spacetime, the wormhole end appears
as a small neighborhood around a closed curve $\cal C$.
As we follow the loop around, we find
that the scalar field winds $n$ times.  If we look at a three-dimensional
slice that intersects the loop, we see magnetic flux coming out;
or if our three-dimensional slice contains the loop, it
changes the magnetic flux, as follows.  Consider the magnetic flux lines in
the background ``time''-slice before and after the slice in which the loop
sits. Since the magnetic flux from the loop changes sign between the ``before''
and ``after'' slices,
one finds that the insertion of the wormhole end creates a loop
of magnetic flux in the background.  Thus the effect of inserting the
wormhole end will be similar to the effect of
insertions of an ``'t~Hooft loop'' operator.\ref\thooft{G.~'t~Hooft,
\nuc{B138}:1 (1978)}
The 't~Hooft loop, $B(\C)$, is the analog in one
extra dimension of the flux creation operator $\phi({\bf x})$ described in
\magworm\ and \thooft.  Its action on a state
({\it i.e.,} a time slice) is to perform a singular gauge transformation
that has a non-zero winding number along a curve that links the loop.  In
the path integral, an insertion of $B(\C)$ means that one integrates over
gauge field configurations such that $\C$ is the world line of a Dirac
monopole singularity of the gauge field.
There is no particular conservation law that forbids the formation of
loops of magnetic flux. But recall that there is an axion winding number
as we follow the loop around: The wormhole insertion creates a loop of flux
with axion winding number. This ``flux winding number'' is
topologically conserved, as follows.

Consider a loop of flux that winds in the manner described above.  The
axion winding number is given by
$$ 2 \pi n  = \oint_{\cal C} {\pmb{\hbox{$ \nabla $}}}\Theta\cdot \d{\bf l}, $$
where the line integral is along the loop.  If one multiplies this by the flux,
one has
$$ 2 \pi n\Phi =  \int {\bf B}\cdot \d {\bf a} \oint_{\cal C}
{\pmb{\hbox{$ \nabla $}}} \Theta \cdot \d{\bf l} =
\int \!\d^3x {\bf B} \cdot {\pmb{\hbox{$ \nabla $}}} \Theta . $$
This is a conserved charge, because
$ {\bf B} \cdot {\pmb{\hbox{$ \nabla $}}} \Theta = J^0 $, where
$$J^\lambda  = \epsilon^{\lambda\mu\nu\rho}F_{\mu\nu}\partial_\rho \Theta ,$$
and $J^\lambda$ is an identically conserved current.  So ``flux winding''
as defined here is a topologically conserved charge.

Unfortunately, it also happens to be zero!  Note that because of flux
conservation, ${\bf B}\cdot{\pmb{\hbox{$\nabla$}}}\Theta =
{\pmb{\hbox{$ \nabla $}}}\cdot
({\bf B}\Theta)$, so our charge is equal to the integral of ${\bf B}\Theta$
over the two-sphere at infinity.  Since there are no monopoles in the
theory, this charge will be zero.  In other words,
if the field $\Theta$ is continuous everywhere, the winding number around
any contractible loop, and therefore {\it any} loop in $R^3$, is zero.
This charge can be made non-trivial by allowing
singularities in $\Theta$.  This is perfectly natural if $\Theta$ is
actually the phase of a complex scalar field $\phi$; the singularities
of $\Theta$ are simply zeros of $\phi$.

\newsec{Superconducting cosmic strings}

One can gain a more complete understanding of the low energy effects of
these wormholes in a theory with vortex solutions.  Consider the
simplest theory with bosonic superconducting cosmic
strings.\ref\witten{E.~Witten, \nuc{B249}:557 (1985)}
This theory has two independent $U(1)$ gauge fields $R_\mu$ and $A_\mu$,
and two complex scalar fields $\sigma$ and $\phi$ that are minimally
coupled to $R$ and $A$, respectively.\foot{The notation here for $A$ and
$R$ is reversed from that of \witten.}  The scalar potential is such that
$\phi$ has a vacuum expectation value, but $\sigma$ does not.  Thus the $A$
gauge symmetry is realized in the Higgs phase, and the $R$ symmetry is
realized in the Coulomb phase.  The scalar potential also has the property
that at the core of a Nielsen-Olesen vortex of the $A$ symmetry, $\sigma$
has a non-zero expectation value.  Since $\sigma$ carries the charge of the
unbroken gauge field $R_\mu$, the $\sigma$ condensate at the core of the
string causes it to be a superconductor.  Loops of this string carry
persistent currents that are characterized by the winding number of the
$\sigma$ field around the loop.  There exist static solutions to the field
equations of this theory known as
``springs''\ref\spring{E.~Copeland, M.~Hindmarsh and N.~Turok, \prl{58}:1910
(1987)\semi
E.~Copeland, D.~Haws, M.~Hindmarsh and N.~Turok, \nuc{B306}:908 (1988)\semi
D.~Haws, M.~Hindmarsh and N.~Turok, \pl{B209}:255 (1988)} or
``vortons,''\ref\vorton{R.~L.~Davis and E.~P.~S.~Shellard, \nuc{B323}:209
(1989)\semi
R.~L.~Davis, \nuc{B325}:125 (1989)}
that consist of a loop of superconducting string prevented from
collapsing by a persistent current and the magnetic field that it generates.

If one identifies the $F_{\mu\nu}$ of our wormhole solution with the field
strength of the $A_\mu$ gauge field and identifies the $\Theta$ field of our
wormhole solution with the phase of $\sigma$, then these vortons are
carriers of exactly the topological charge defined.  The theories
do not match exactly, but in the limit where the radius of the
two-sphere boundary of the wormhole is much less than the radius of the
core of the string, one might expect that the wormhole solution can
successfully patch on to the vorton solution.
In the theory with superconducting cosmic strings, of course, the
periodic scalar is coupled to a gauge field.  Adding this gauge field to
the theory explicitly may shed some additional light on the dynamics of
the wormhole in this background.

The new theory has the action
$$ \eqalign{
  S = \intst \biggl[ -{1 \over 16 \pi G} R &+ {v^2 \over 2} g^{\mu\nu}
       (\partial_\mu \Theta + R_\mu)(\partial_\nu \Theta + R_\nu) \cr
 &+ {1\over 4 e^2} g^{\mu\nu} g^{\lambda\sigma} F_{\mu\lambda} F_{\nu\sigma} +
{1\over 4 e'^2} g^{\mu\nu} g^{\lambda\sigma} G_{\mu\lambda} G_{\nu\sigma}
 \biggr], \cr } $$
where $R_\mu$ is the gauge field that couples to $\Theta$,
$G_{\mu\nu}$ is the field strength for $R$, and $e'$ is the coupling.
In this model, the field equations for $F$ are unchanged. The field equation
for $\Theta$ becomes:
$$ \partial_\mu [ \sqrt{g} g^{\mu\nu}( \partial_\nu \Theta + R_\mu) ] = 0. $$
The $G$ field equation is
$$ \partial_\mu [\sqrt{g} G^{\mu\nu} ] = (e' v)^2 \sqrt{g}
(\partial^\nu \Theta + R^\nu) . $$
The energy-momentum tensor (and thus the Einstein equations)
will be given by adding a term for $G$, and by
replacing $\partial_\mu \Theta$ with $\partial_\mu \Theta + R_\mu$
throughout.

Now consider the wormhole ansatz.  The situation can be simplified somewhat
by setting $\Theta$ to zero identically via a gauge
transformation.  Then any non-zero winding number for $\Theta$ around the
$S^1$ becomes a non-zero Wilson loop for $R^\mu$.  This will not, however,
be constrained to be an integer.  The ansatz for $R$ will be
$$ R_\mu = 2 \pi n f(\tau) \delta^l_\mu, $$
that is, only $R_l$ is non-zero and it only depends on $\tau$.  With this
ansatz, the $\Theta$ field equation is satisfied automatically.
The field equation for $G$ becomes
\def\fd{{\dot f}}
\eqn\feq{ {\ddot f} = \left( {\ad \over a} - {2 \bd \over b} \right)
          {\dot f} + m^2 f . }
The Einstein equations become
\eqna\einsnew
$$ \eqalignno{
   {2 \ad \bd \over a b} + {\bd^2 \over b^2} - {1\over b^2} &=
   -{Q_1^2\over a^2} \left( f^2 - {\fd^2 \over m^2} \right)
     - {Q_2^2\over b^4}, & \einsnew a \cr
   {2 \bdd \over b} + {\bd^2 \over b^2} - {1\over b^2} &=
   {Q_1^2\over a^2} \left( f^2 + {\fd^2 \over m^2} \right)
     - {Q_2^2\over b^4}, & \einsnew b \cr
   {\add \over a} + {\bdd \over b} + {\ad \bd \over a b} &=
   -{Q_1^2\over a^2} \left( f^2 + {\fd^2 \over m^2} \right)
     + {Q_2^2\over b^4}, & \einsnew c \cr} $$
where I define $m = e' v$ and define the other quantities as before.

The conditions at the wormhole throat will be mostly unchanged.  One
still wants $\ad = \bd = 0$ and $\add > 0, \bdd > 0$.  So that the
throat will be time-symmetric, I impose the additional requirement that
$\fd = 0$.  Without loss of generality, I set $f_0 = 1$,
since one can always rescale $Q_1$ to compensate.\foot{I chose
the normalization of $f$ so that in the $m \to 0$ limit, $f(\tau) = 1$ gives
solutions equivalent to the ones previously found.}
In this case, the throat conditions \throat\ and \thrin\ will
both hold unchanged.  The scaling properties given by \scaling\ will
also hold, provided one scales $m$ as $m \to \lambda_2 m$.  In other
words, the $Q_1 = Q_2 = 1$ solutions are general, provided one holds
$b_0/Q_2$ and $m/Q_2$ fixed as one varies $Q_1$ and $Q_2$.

I can now integrate the equations.  I find that when $m$ is small, the
solutions are not affected over timescales of order $m^{-1}$.  When $m$
is large, however, $f$ grows rapidly and leads to collapse of the
wormhole.  An intermediate situation is shown in
Figure~4.

I also attempt to further approximate the core of a superconducting
cosmic string by letting the expectation value of the scalar field go to
zero after some distance ({\it i.e.}\ outside the ``core'').  In this case
the solution asymptotically approaches a flat, $a \to
\hbox{constant}$, $b \to \tau + \hbox{constant} $ solution.  Indeed, if
the expectation value goes to zero in a ``nearly flat''
regime before the collapse has
set in, the relaxation occurs fairly rapidly.  In such a region, we can
approximate equation \feq\ as
$$ {\ddot f} \simeq {-2 \over \tau} \fd , $$
for which $ \fd \sim \tau^{-2}$; thus $f$ relaxes rapidly to a constant,
allowing the geometry to be asymptotically flat.
This simple calculation shows that putting the solution in a
slightly more physical context can improve the matching to the
background geometry, and can increase one's confidence that the
hypothesis (that exact solutions of this approximate form exist) is
correct.

Insertion of this wormhole solution will
induce an operator that creates or destroys a vorton. One can visualize
this in the following way:  Imagine a loop of superconducting cosmic
string propagating forward in time, leaving a world tube.  A spacelike
hypersurface intersects this tube in a loop, and it is along this loop
(in the core of the vorton) that we insert the wormhole end.
The vorton core has magnetic flux flowing along it, and has a winding
number for the scalar field which is non-zero in the core.
As discussed before, the wormhole
insertion destroys (or creates) a unit of magnetic flux whenever the
scalar field winding number is non-zero.  This reduces (or increases)
the flux-winding number product which is the conserved charge of the
vorton.  Since the
vorton is quantum-mechanically unstable, the charge is not strictly
conserved, but if the vorton lifetime is sufficiently long, the wormhole
may be the most important contribution to its decay.  (One would expect
this to be the case when the additional action of inserting the
wormhole solution is less than the additional action of inserting an
instanton in the world sheet of the string that allows the winding
number of the scalar field along the string to decrease.  Note that this
decay, which changes the winding number, is not quite the same as the
wormhole-mediated decay, which changes the flux.)
The wormhole
contribution to the theory could then be described in terms of an effective
local field that describes the vorton degrees of freedom, even though this
somewhat obscures the non-local nature of the wormhole insertions.

Let me now ask which configurations
dominate the path integral.  I have found solutions
which, for a fixed value of the charges $Q_1$ and $Q_2$, can be
patched to the background spacetime along an arbitrarily long curve
${\cal C}$ by adjusting the parameter $Q_2^2/b_0^2$ at the wormhole throat.
In searching for the lowest action classical solution contributing to a
given process, however, one cannot fix the parameters at the throat --- one
can only fix parameters on the background.  For example, if one is looking
for the leading contribution to a process which carries away a fixed
magnetic flux with a fixed scalar winding number, one should include only
the lowest action solution for fixed $Q_1$ and $Q_2$.  I expect that this
will be a circular wormhole solution of some fixed length; this is
analogous to the stable static vorton solution with similar parameters.  One
could also imagine actually searching for processes which annihilate a
vorton of a given size as well as charge, the dominant contribution to which
will be given by a wormhole of the appropriate size.  The correct wormhole
solution always depends on the amplitude under consideration.

\newsec{Conclusions}

I have constructed wormhole solutions of topology $S^2 \times S^1 \times
R$ in a theory with electromagnetic fields and periodic scalar fields,
as well as in a gauged version of this theory.
While not exactly realistic, this theory is a simple example of a theory
with topologically conserved charges on both the two-sphere and the circle.
These wormhole solutions do not fit the paradigm of having $S^3\times R$
topology and asymptotic flatness in both directions.

These wormholes may be sensibly interpreted
in terms of effective 't~Hooft loop operators (or monopole loops) on the
background spacetime, but this interpretation still leaves a number of
loose ends.
For example, it is an unproven hypothesis that the asymptotically flat
solutions suggested by this work actually exist.  Even if they do, they may
not contribute to the Euclidean path integral in the simple form suggested.

Another issue concerns the Coleman-Lee solution to the large wormhole
problem.\ref\escape{S.~Coleman and K.~Lee, \pl{B221}:242 (1989)}
Although these wormholes are indeed supported by a
conserved charge, it is not obvious how this charge can be ``drained away''
by smaller wormholes.  This might be easiest to see in terms
of a field operator that creates or destroys vortons.  But in this case,
the global, topological aspects of the conserved charge seem to be lost.

Despite these difficulties, it seems quite likely that there are in fact
solutions to the Euclidean Einstein equations that connect an
asymptotically flat background to an $S^1 \times S^2$ baby universe.
These wormhole solutions will give the dominant contributions
to violation of vorton charge conservation, and thus must be considered
on an equal basis with other wormhole contributions to low-energy
physics.  Thus any complete description of low-energy physics that
includes wormholes must be able to reckon with solutions of the type I
have constructed here.

\newsec{Acknowledgments}

I would like to thank John Preskill for suggesting this line of research
and for helpful discussions along the way.  I also had helpful
discussions with Erick Weinberg and Kimyeong Lee.

\listrefs
\ifig\figA{$a(\tau)$ (solid) and $b(\tau)$ (dashed) for $Q_1 =
Q_2 = 1$ and $Q_2^2/b_0^2 = 0.98$.}{\epsfbox[0 0 288 240]{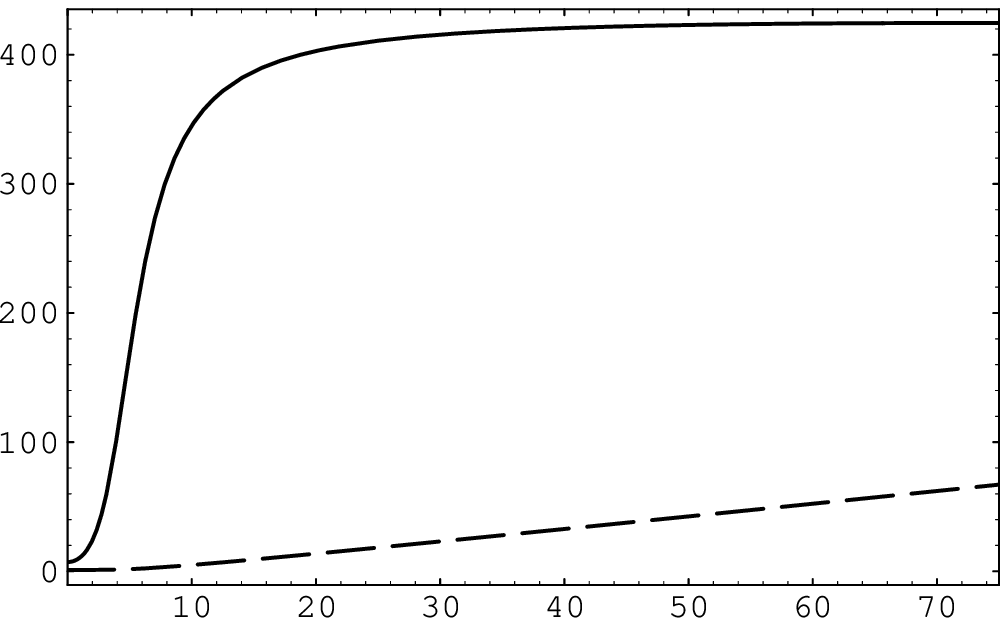}}
\ifig\figB{$a(\tau)$ (solid) and $b(\tau)$ (dashed) for $Q_1 =
Q_2 = 1$ and $Q_2^2/b_0^2 = 0.8$.  The figure does not show it,
but $b$ goes to infinity where $a$ goes to
zero.}{\epsfbox[0 0 288 240]{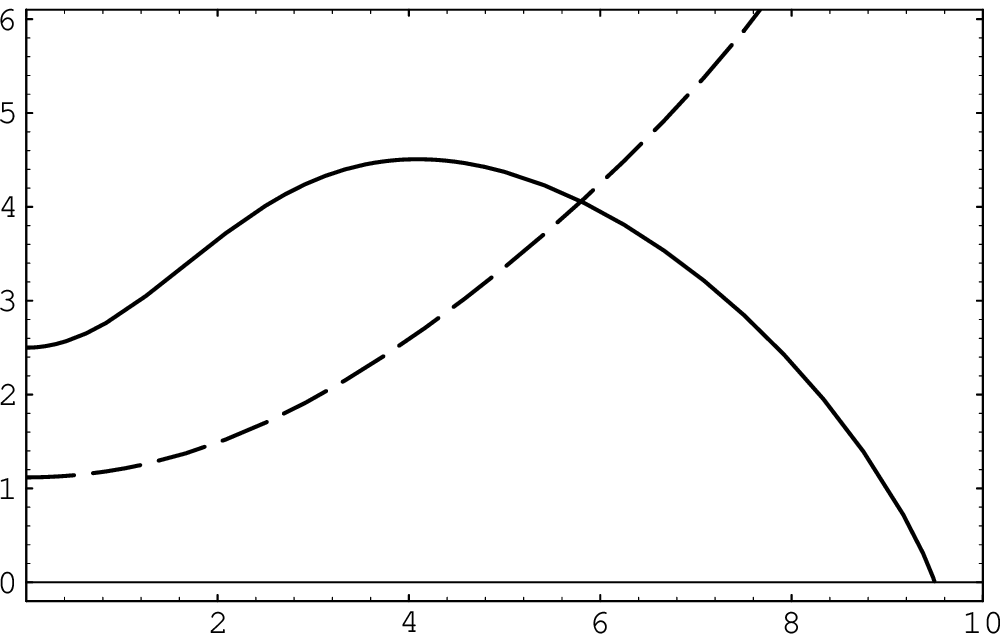}}
\ifig\figC{$a(\tau)$ (top) and $b(\tau)$ (bottom) for $Q_1 =
Q_2 = 1$ and $Q_2^2/b_0^2 = 0.999$.  The maximum value of $a$ increases
dramatically as
$Q_2^2/b_0^2 \to 1$.}%
{\vbox{\epsfbox[0 0 288 240]{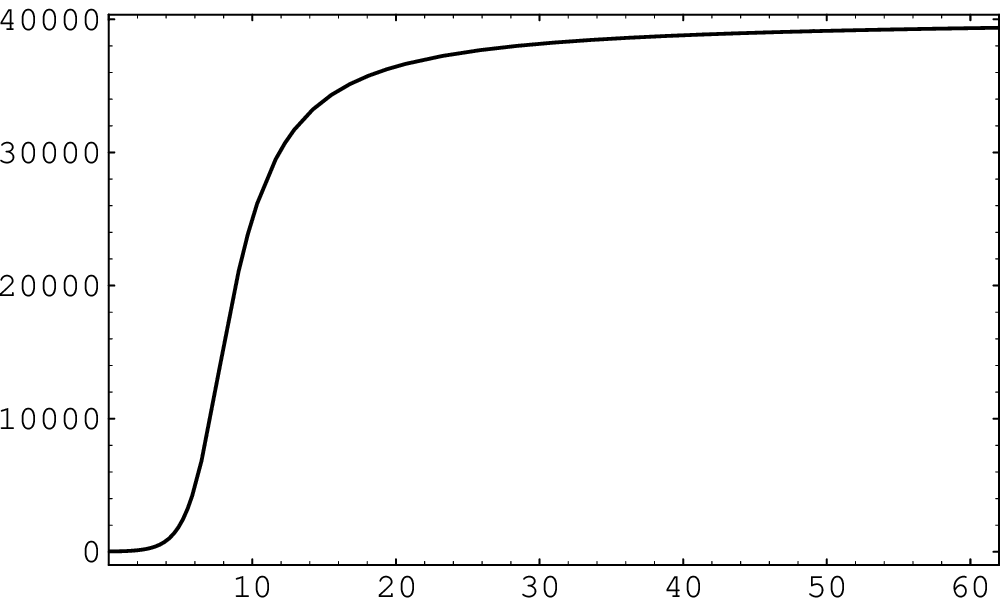}\epsfbox[0 0 288 240]{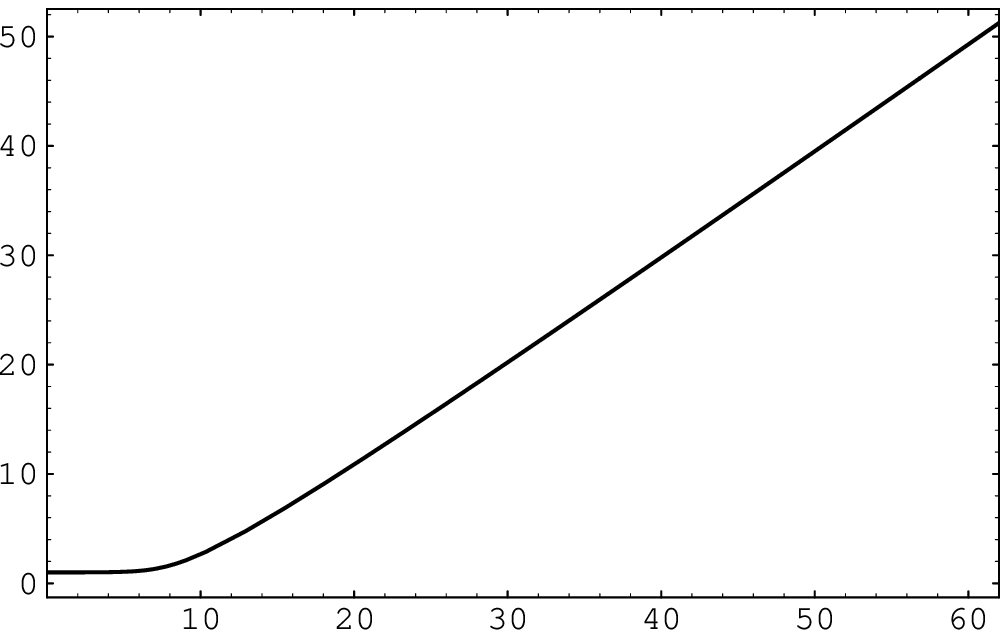}}}
\ifig\figD{The top figure shows $a(\tau)$ (solid) and $b(\tau)$ (dashed)
for $Q_1 = Q_2 = 1$,  $Q_2^2/b_0^2 = 0.98$, and $m = 0.1$.  The bottom
figure shows $f(\tau)$ for these
parameters.}%
{\vbox{\epsfbox[0 0 288 240]{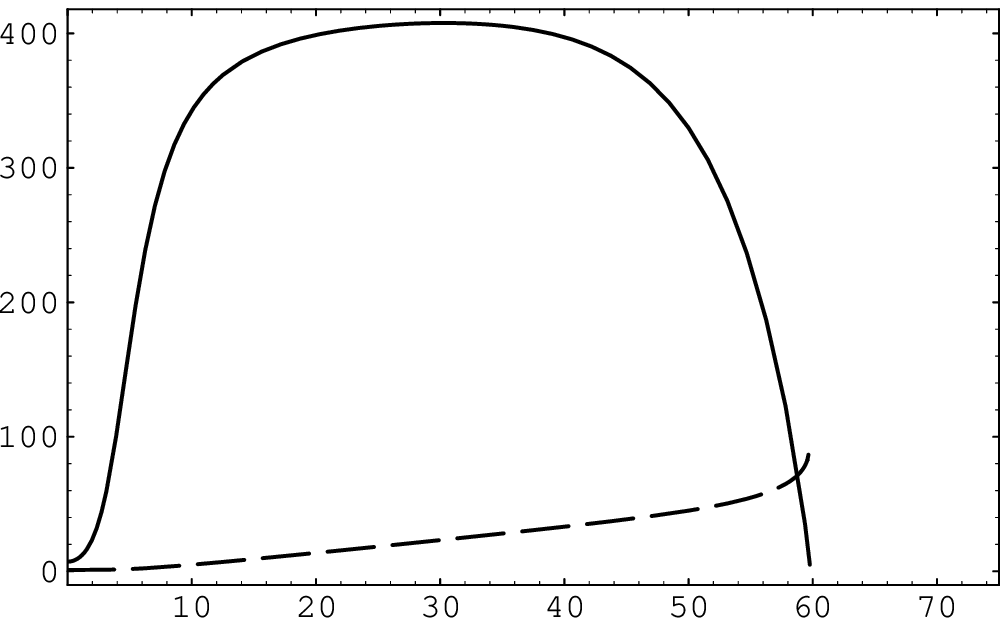}\epsfbox[0 0 288 240]{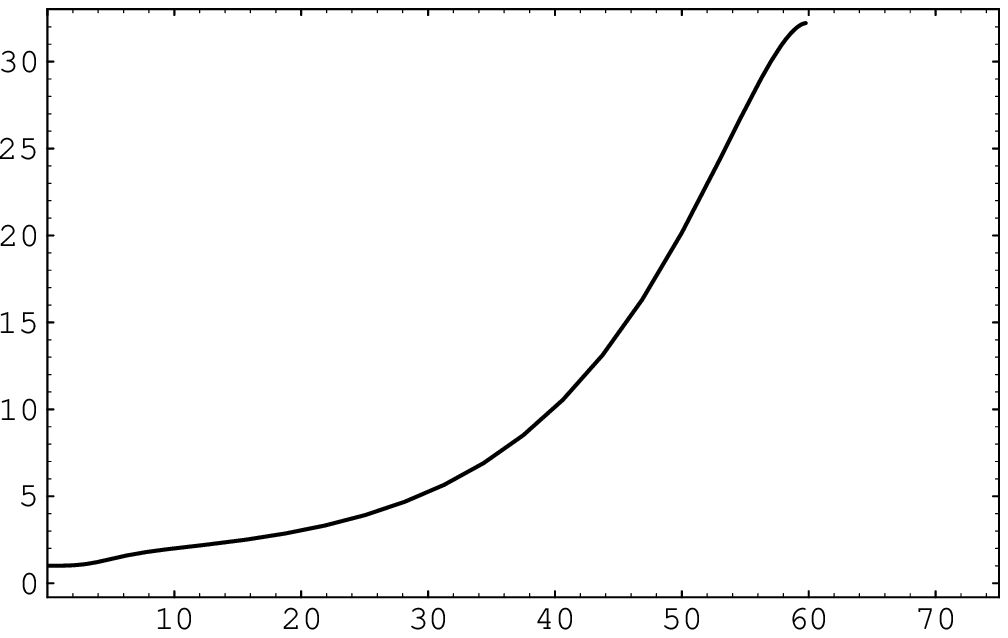}}}

\bye